\documentclass[11pt]{article}
\usepackage{apalike,latexsym,amssymb,bbm}

\def\titel{Constraints on Covariance}

\def\ben{\begin{enumerate}}
\def\een{\end{enumerate}}
\def\bi{\begin{itemize}}
\def\ei{\end{itemize}}
\def\bq{\begin{quote}}
\def\eq{\end{quote}}

\def\be{\begin{equation}}
\def\ee{\end{equation}}
\def\ba{\begin{array}}
\def\ea{\end{array}}

\begin{document}
\sloppy

\thispagestyle{empty}

\vspace*{-24mm}

\vspace{15mm}

{\Large\bf \titel}

\vspace{7mm}

\hfill\parbox{13cm}{
{\sc Bing Hurl}

\vspace{4mm}

{\small Institute of High Energy Physics, P.O.Box 918-4,
Beijing 100039,
China\\[1mm]
\hspace*{3mm}\\[1mm]
e-mail: heb@alpha02.ihep.ac.cn.

\vspace{7mm}

\today

\vspace{4mm}

\paragraph{Abstract.}Based on the principle of relativity, we find that the
sufficient and necessary condition for the general covariance of a field theory
actually requires more than the invariance of its local Langrangian density. 
If the spacetime is not a flat one, its derivative requirement from the analysis of
the parallel transportation of tensor fields over spacetime restricts the generally
supposed covariance group, the group of differmorphisms, to the group of linear
coordinate transformations. Moreover, for any of a field theory with linear
equations of motion, it stipulates for a universal physical propagation speed 
of interaction over the spacetime manifold. 
}
}\hspace*{5mm}

\vspace{9mm}


\section{Introduction}
General covariance, as the technical realization of the principle of
general relativity, is regarded as one of the most powerful symmetries
of nature. Its conceptual simplicity leads many to believe that
it should be a prominent attribute of all fundamental physical laws.
A physical theory is said to be covariant between two systems of coordinates,
say K and K', if all the concerned physical quantities 
(including
the observables) expressed or measured in them are in one-to-one
correspondence according to the respective laws of tensor and spinor
transformations induced by the spacetime coordinate transformation connecting
K and K' and, as a result, the described physical processes 
take place in
the similar manner in both of them if the transformation also keeps the
metric field form-invariant. Under this view, there is general covariance if 
this property holds good to an arbitrary (differentiable)
spacetime coordinate transformation
\begin{eqnarray}
x^{\mu'}=f^{\mu}(x^{\nu}).   
\end{eqnarray}
For a field action on a general spacetime manifold M with metric field
$g_{\mu\nu}(x)$ , e.g. the action
\begin{eqnarray}
S=\int\limits_{M} \sqrt{|g(x)|}d^4 xL(x) =\int\limits_{M}\sqrt{|g(x)|}d^4 x
\left[- \frac{1}{4}(\partial_{\mu}A_{\nu}-\partial_{\nu}A_{\mu})
(\partial^{\mu}A^{\nu}-\partial^{\nu}A^{\mu})
-\frac{1}{c}J_{\mu}A^{\mu}\right]
\end{eqnarray}
that describes an Abelian gauge field with a linear coupling to
the external source
field $J^{\mu}(x)$, the following transformations
\begin{eqnarray}
A^{\mu'}(x')=\frac{\partial x^{\mu'}}{\partial x^{\mu}}A^{\mu}(x),
\end{eqnarray}
\begin{eqnarray}
J^{\mu'}(x')=\frac{\partial x^{\mu'}}{\partial x^{\mu}}J^{\mu}(x)
\end{eqnarray}
\begin{eqnarray}
g_{\mu'\nu'}(x')=\frac{\partial x^{\mu}}{\partial x^{\mu'}}
\frac{\partial x^{\nu}}{\partial x^{\nu'}}g_{\mu\nu}(x)
\end{eqnarray}
induced by the arbitrary spacetime transformation (1) leave the field action $S$
invariant and
therefore, through the variation with respect to $A^{\mu}(x)$,
the local equations of motion take on the same form in all systems of
coordinates.  In
this way the related
physical processes involving $A^{\mu}(x)$ and $J^{\mu}(x)$ are 
thought to take place similarly in different systems of coordinate, 
given that the initial, boundary
data for the field equations should be the same and the metric field should be
form-invariant in these systems. The invariance of the field action or 
the invariance
of Lagrangian density under the spacetime coordinate transformation is
presently used as the sufficient condition for the covariance
of any field theory, be its equations of motion linear or nonlinear.

It has been long recognized that general covariance for a physical theory is a
mathematical rather than a physical requirement \cite{K17} \cite{F59} \cite{M75}
\cite{F83}; nevertheless,
it stipulates for the way in which the local quantities should be 
mathematically
correlated, such as those in Eqs. (3)-(5). In the more recent literature
general covariance group is identified with the group of differmorphisms of
spacetime manifolds on which the fundamental physical laws can be reduced
to equations involving only local differential operators of finite order.
Various theories involving gravitational effect have this group as the largest
one
preserving the form-invariance of local field equations, while specify some
subgroup of it as the group of automorphisms to preserve certain 
geometrical structures,
called absolute elements in \cite{T66}, \cite{A67}, \cite{T73}, invariant.
The local quantities in a physical theory must be in one-to-one
correspondence according to the laws of tensor and spinor transformations, as 
long as the theory is a covariant one; once the measures or expressions of
them are determined  
in one coordinate system, those in all other 
physically admissible coordinates are also determined if the concerned spacetime
transformations are specified. Obviously it is the consequence of the validity
of the principle of
general relativity or the principle of general invariance, which requires that
all the admissible systems of coordinate be equal in the description of nature.

With a closer look at the original meaning of the principle of relativity,
however, we find that there is a stronger condition on the covariance of
a field theory established on general spacetime manifold, i.e. that for the 
consistency of description of the physical
processes in different coordinate systems, and it limits the arbitrary 
spacetime transformation group (1) 
to its much
smaller subgroup, which is picked out to meet the requirement for consistency.
Hereafter, for a distinction from `physically covariant' 
or `covariant', we
will call the field 
equations of a field action `form covariant' if
its Lagrangian density satisfies the invariance under spacetime coordinate
transformation. 
In this letter we first present a brief discussion on
our sufficient and necessary condition for the covariance of a field theory,
then apply it to the study of the covariance of the field described by
Eq. (2) under arbitrary spacetime transformation (1), and finally we use
it to study the covariance of the field theory under linear transformations in
homogeous spacetime. The advantage of choosing action (2) as the example is
that the linearity of its field equations allows us to have a better
understanding of their solution structure.

\section{ Sufficient and necessary condition for covariance}
The equivalence
of systems of coordinate in describing a covariant field theory is the
essential
requirement raised by the principle of relativity. Here the term 
`equivalence' has a two-fold meaning: 1) The  
definite correlation of the local physical
quantities in terms of
the transformations of tensors (spinors) between any couple of systems
of coordinate in the equivalent class; 2) The arbitrariness in 
selecting a coordinate among the
equivalent ones for the construction
of solution to these field equations. Only after the two points are met,
can system K with the
geometrical structure in it described by $g_{\mu \nu}(x)$, 
$\Gamma ^{\tau}_{\mu
\nu}(x)$,
etc. of M, be exactly equivalent to system K', where
the geometrical structure is characterized by $g_{\mu' \nu'}(x')$,
$\Gamma^{\tau'}_{\mu' \nu'}(x')$, etc. of M' (the differmorphic image
of M), and the field theory under study respectively by the observors
in K and K' be truly covariant.
Of course the realization of the completely equivalent description of the
fields (at the classical level) among different systems of
coodinate also requires that 
the boundary data and initial data for the field equations should be 
preserved in the proper way
under the concerned transformations, but here we only consider 
the boundary and time origin
at infinity and therefore their influence are
negligible. 

For clarification we construct the following diagram to illustrate the
equivalence of system K and K' in the study of a covariant field theory (here
the
action (2) is taken as example).
\be
\ba{rcccl}
 & g_{\mu\nu}(x), J^{\mu}(x) & \stackrel{}{\longrightarrow} &
g_{\mu'\nu'}(x'), J^{\mu'}(x') & \\
 {} & \downarrow &     & \downarrow & {} \\
 & A^{\mu}(x)  & \stackrel{}{\longrightarrow} & A^{\mu'}(x')
\ea
\ee
In this diagram the horizontal arrows represent the transformations of the related
fields from system K to K', and the perpendicular arrows the construction of
the solution to field equations with the source field and geometrical
structure in K and K', respectively.
Thus the equivalence of K and K' leads us to the sufficient and necessary
condition
for the covariance of a field theory: a field theory is covariant
between two systems of coordinate, if and only
if such kind of diagram for it commutes. The `only if' part of the condition
is more nontrivial; it guarantees the compatibility of the transformations
for the geometrical structure of the spacetime manifold with those for
the physical processes taking
place on the spacetime manifold.

For a better understanding of the condition we imagine such a
picture:
In system K established on M, the physical fields $A^{\mu}(x), J^{\mu}(x)$
in action (2) are determined by an observor through measurement (and
solving the field equations) and, after the
specification of the exact form of spacetime coordinates transformation
between K and K',
they are transformed according to Eqs. (3) and (4) to $A_{1}^{\mu'}(x'),
J^{\mu'}(x')$
in K', where another observor
solves the form covariant field equations only with the transformed
$ J^{\mu'}(x')$ and
$g_{\mu'\nu'}(x')$. Unless the diagram is commutative, the solution
$A_{2}^{\mu'}(x')$ thus constructed in K', which is compatible with the 
geometrical
struture described by $g_{\mu'\nu'}(x')$ there, will not be equal to
$A_{1}^{\mu'}(x')$, and therefore an awkward situation will arise from the
ambiguity of $A^{\mu'}(x')$. At first sight this requirement seems redundant since
any form covariant equation is expressed in the form:
 a tensor (spinor) = 0,
which is true to any coordinate system, and indicates that $A_{1}^{\mu'}(x')$ 
transformed from
$A^{\mu}(x)$ according to Eq. (3) should satisfy the field equations in K' too.
However, we will see later that the undesired noncommutation of the above
diagram does arise from the non-local geometrical structure we have to take
into account in the construction of a solution involving the propagation of
physical fields from one point on spacetime manifold to another, which is
governed by the natural geometry of the spacetime \cite [p.173]{L90}.

\section{ Application to an arbitrary spacetime coordinate transformation}
For simplicity we study $A^{\mu}(x)$ produced by a point particle 
moving in the region of an arbitrary spacetime manifold without
the presence
of other matter and radiation. If the radiation of its own is week enough,
it can be regarded as moving over the spacetime backgroud with
almost identically
vanishing Ricci tesor field $R_{\mu\nu}$ (from
field equations $R_{\mu\nu}=\frac{8\pi
G_N}{c^4} (T_{\mu\nu}-\frac{1}{2} g_{\mu\nu}T)$ ).
The source field in this case can be expressed as \cite[p.165]{B80}
\cite[p.252]{S84}
\begin{eqnarray}
J^{\mu}(x)=\frac{ec}{\sqrt{|g(x)|}}\int
\limits_{-\infty}^{+\infty}ds\frac{dx_{0}(s)^{\mu}}{ds}\delta(x-x_{0}(s)),
\end{eqnarray}
where $dx_{0}(s)^{\mu}/ds$ is the four-velocity of the particle, s the
proper time, and $ec$ the coupling constant.
Throughout our discussion Lorentz gauge (in the sense of covariant
derivative) may be used. The linearity of the equations of motion derived 
from action (2) determines
definitely a light cone structure (a hypersurface on spacetime manifold) 
at each point on M. The point particle
in Eq. (6) contributes only to field $A^{\mu}(x)$ over and inside 
its forward light cone
(the propagation of $A^{\mu}(x)$ along the surface of light cone and its
scattering by the interaction with curvature of spacetime to the inside of
the light cone \cite[p.72]{St82}),
because of the retardation effect caused by the limited propagtion speed
of interaction. Although the exact equations of light cone on a general
spacetime manifold is hard to know, we are sure of their existence
by constructing the local light cones with locally constant $g_{\mu \nu}(x)$ 
( in the form of local world line element ) and
gluing them together to form
a global one. More generally, if the source field $J^{\mu}(x)$ is
distributed on
the whole spacetime manifold, the contribution to $A^{\mu}(x)$
comes from the integral inside and over the backward light cone of point $x$.

On the subset of a spacetime manifold where  
$R_{\mu\nu}\approx 0$, 
the physical considerations require that the
propagation of $A^{\mu}(x)$ from the source particle to another spacetime
point 
should be along
the null geodesics connecting them. In a specific system of reference,
say K, an observor sees that the field $A^{\mu}(x)$ at the intersecting
point,
$x^{\mu}$, of his/her world line with 
the forward light cone of an arbitrary point on the world line of 
the source particle
($x^{\mu}_{0}(s)$) is proportional to vector $n^{\mu}(x)$ that is produced
by the parallel translation of the four-velocity vector
$$n^{\mu}(x_{0})=\left(dx_0^{\mu}/ds\right)/\left(
\sqrt{g_{\mu\nu}(x_0)(dx_0^{\mu}/ds)(dx_0^{\nu}/ds)}\right)=dx_0^{\mu}/ds$$ 
along the null geodesics connecting the two points:
\begin{eqnarray}
A^{\mu}(x)=C(x,x_{0})n^{\mu}(x) = C(x,x_{0})\left[n^{\mu}(x_{0})-\int\limits_{x_{0}}^{x}
\Gamma_{\nu\tau}^{\mu}(z)n^{\nu}(z)dz^{\tau}\right],
\end{eqnarray}
where the integral is along the null geodesic, and $C(x,x_{0})$ is a scalar
function which reduces to
\begin{eqnarray}
\frac{e}{4\pi}
\frac{\sqrt{\eta _{\mu
\nu}\left(dx_0^{\mu}/ds\right)\left(dx_0^{\nu}/ds\right)}}
{(x-x_0)^{\tau}(dx_0/ds)_{\tau}}
=\frac{e}{4\pi}\frac{1}{(x-x_0)^{\tau}(dx_0/ds)_{\tau}}
\end{eqnarray}
in Minkowski spacetime. 
One component of the four-velocity of the source particle contributes
to the others of $A^{\mu}(x)$ after the parallel displacement along
the null geodesics, since the field equations on a general spacetime manifold
cannot be split into the uncoupled ones with respect to the components of the
vector fields.
Particularly,
if there is more than one geodesic connnecting the two points ( e.g.
they are conjugate points of each other), $n^{\mu}(x)$ is a sum of the
parallel translations of $n^{\mu}(x_{0})$ along all geodesics.

In the light of the above result, the covariance of the
produced field $A^{\mu}(x)$ on the forward light cone of its source particle
can be studied through the covariance of the field $n^{\mu}(x)$ produced by
the parallel transportation of $n^{\mu}(x_{0})$ along the null geodesics
on the forward light cone of $x_{0}(s)$; the covariance of $n^{\mu}(x)$ is the
necessary condition for the covariance of $A^{\mu}(x)$ on the forward
light cone of its source particle. 
The necessary condition
for the covariance of the field theory described by Eq. (2) is then specified
to the commutation of the following diagram:
\be
\ba{rcccl}
 & n^{\mu}(x_0) & \stackrel{}{\longrightarrow} & n^{\mu'}(x'_0) & \\
 {} & \downarrow &     & \downarrow & {} \\
 & n^{\mu}(x) & \stackrel{}{\longrightarrow} & n^{\mu'}(x')
\ea
\ee
Here the perpendicular arrows represent the parallel displacement of
vectors along the null geodesics to the forward lightcone of $x^{\mu}_{0}$
(resp. $x^{\mu'}_{0}$) in K (resp. K'). 
If we require that $g_{\mu \nu}(x)$ transform as what is given in the first
diagram, the null geodesics connecting
$x^{\mu}$ and $x^{\mu}_{0}$ are always transformed piecewise through
the transformation of the local light cone to those connecting
$x^{\mu'}$ and $x^{\mu'}_{0}$, i.e., 
a null geodesic is generally covariant.
The commutation of the diagram means that the field $n^{\mu}(x)$ produced 
by the parallel displacement of
$n^{\mu}(x_{0})$ along the above-mentioned null geodesics will be 
transformed to the field $n^{\mu'}(x')$ by the parallel
displacement of $n^{\mu'}(x'_{0})$ along the corresponding null geodesics
in K', so the two following results
\begin{eqnarray}
n_1^{\mu'}(x')=\left(\frac{\partial x^{\mu'}}{\partial
x^{\mu}}\right)~n^{\mu}(x)=
\left(\frac{\partial x^{\mu'}}{\partial
x^{\mu}}\right)~(n^{\mu}(x_0)-\int\limits_{x_{0}}^{x}
\Gamma_{\nu\tau}^{\mu}(z)n^{\nu}(z)dz^{\tau}),
\end{eqnarray}
and
\begin{eqnarray}
n_2^{\mu'}(x')&=&n^{\mu'}(x'_0)-\int\limits_{x'_{0}}^{x'}\Gamma
_{\nu'\tau'}^{\mu'}(z')~n^{\nu'}(z')dz^{\tau'}\\ 
&=&\left(\frac{\partial x_0^{\mu'}}{\partial x_0^{\mu}}\right)n^{\mu}(x_0)-
\int\limits_{x_{0}}^{x}\left(\frac{\partial z^{\mu'}}{\partial z^{\mu}}\right)
~\Gamma_{\nu\tau}^{\mu}(z)n^{\nu}(z)dz^{\tau}
+\int\limits_{x_{0}}^{x}\left(\frac{\partial^2 z^{\mu'}}{\partial z^{\nu}
\partial z^{\tau}}\right)n^{\nu}(z)dz^{\tau}     
\end{eqnarray}
must be equal.

Two different cases we should consider: 1) The curvature tensor field $R^{\mu}_{\nu\tau\sigma}(x)$ is non-vanishing in K.
Obviously, the two results are not equal for a nonlinear transformation,
with
the different transformation Jacobians at different points
and an additional summand in Eq. (11) 
\footnote[1]{ The transformation of
$\Gamma_{\nu\tau}^{\mu}$ in Eq. (11) is defined to guarantee the commutation 
of the second
diagram under arbitrary spacetime transformation (1) in a small neighborhood of $x_{0}$
(see \cite[p.276]{M52}. But it
depends on considering the Taylor expansion of the transformation
Jacobian (at $x$)
around $x_{0}$ only to the first order and the fact
$dx^{\mu'}=\frac{\partial x^{\mu'}}{\partial x^{\mu}}dx^{\mu}$,
which are invalid for a finite distance from $x_{0}$. }.
Thus, on a general spacetime maniold, the only candidates
that preserve the covariance of the field theory
descibed by Eq. (2) are linear transformation groups.
2) The curvature tensor field $R^{\mu}_{\nu\tau\sigma}(x)$
vanishes identically in K.
Because $n^{\mu}$ in this case remains constant under parallel
displacement along any of a curve in K, i.e.
the solution of the partial differntial equation,
\begin{eqnarray}
\partial_{\mu}n^{\nu}(x)=-\Gamma^{\nu}_{\mu\tau}(x)n^{\tau}(x),
\end{eqnarray}
exists \cite[p.167]{B47}, $n_{1}^{\mu'}(x')$ and
$n_{2}^{\mu'}(x')$ will be absolutely equal under an arbitrary spacetime transformation (1).

For completeness we can add the gravitational field term
\begin{eqnarray}
L_{G}=\frac{c^4}{16\pi G_N} g_{\mu\nu}R^{\mu\nu} 
\end{eqnarray}
with
\begin{eqnarray}
R_{\mu\nu}=\partial_{\alpha}\Gamma_{\mu\nu}^{\alpha}
-\partial_{\nu}\Gamma_{\mu\alpha}^{\alpha}
+\Gamma_{\alpha\beta}^{\beta}\Gamma_{\mu\nu}^{\alpha}
-\Gamma_{\nu\alpha}^{\beta}\Gamma_{\mu\beta}^{\alpha}
\end{eqnarray} 
to the Lagrangian density in Eq. (2).
Because of the presence of the nonlinear self-interaction term 
the
structure of the solution to the field equations is beyond
our knowledge, so, except for a few special cases ( e.g. the propagation of
gravitation wave discussed in terms of its wave front equation 
in a harmonic
coordinate system given in \cite[p.175]{F59}, we are not sure whether or not 
gravitational
wave propages along the null geodesics on the forward light cone
of its source.
A linear approximation of the theory can however be
realized if we consider only the gravitational perturbation
$\epsilon^{\mu\nu}(x) $
over the spacetime background $h^{\mu \nu}(x)$:
\begin{eqnarray}
g^{\mu\nu}(x)=h^{\mu\nu}(x)+\epsilon^{\mu\nu}(x),   
\end{eqnarray}
then the covariance of $\epsilon^{\mu\nu}(x) $ can be discussed
in the similar way. Together with the our previous discussion
on the covariance of $A^{\mu}(x)$ we conclude that the form
covariance of the field equations is not sufficient to guarantee
the physical covariance of the fields. In other words, it doesn't
generally
enable the physical fields (classical solutions to the actions of
the fundamental interactions) expressed or measured in different
coordinate systems to be in one-to-one correspondence through
the tensor transformations induced by the arbitrary spacetime
transformation (1) connecting them, since the parallel transportation
of tensor fields should be an absolute element in a field theory.

The fact that the physically admissible spacetime coordinate transformations
in a curved spacetime (in the sense of maintenance of physical covariance) are
only those of
linear transformation groups implies the existence of some preferred
systems for the study of the physical phenomena in large spacetime scale.
For a possible realization of physical covariance under the arbitrary spacetime
coordinate transformation (1), the structure of spacetime
in which all physical processes take place
must be that of an absolutely flat one---the curvature tensor field
$R^{\mu}_{\rho \nu \sigma}(x)$
degenerates identically over the spacetime manifold.
So physical covariance under the arbitrary spacetime coordinate transformation (1) 
is connected with the
flatness of spacetime (in curved spacetime 
the covariance of the
field theories under arbitrary spacetime coordinate transformation (1) 
in a small neighborhood of every spacetime point can be regarded as the
result of the local
Minkowskian property from the principle of equivalence), 
and it is necessary, as it
was originally pointed out by V. A. Fock \cite[p.370]{F59}, to distinguish
relativity in the sense of uniformity of spacetime from relativity
in the sense of the possibility of using arbitrary reference
systems. In the theory of relativity with the latter sense, the 
term `motion' is defined by the
Jacobian matrix $\left(\partial x^{\mu'}/\partial
x^{\mu}\right) $
between two systems of reference \cite[p.7]{S82}.
As a matter of fact, however, the mathematical expression for a
spacetime coordinate transformation between two arbitrary
systems of reference established on general spacetime manifolds can never be
determined. In one of the
systems an observor only knows (through measurement) the trajectory of the 
other
expreesed by $x^{i}(t)$ (i=1,2,3), where $t$ is the coordinate time.
The additional symmetry, the existence of which 
entails certain degree of homogeneity for spacetime structure, 
is indispensible 
for the specification of
the exact form of spacetime transformation between two reference
systems. For example, the establishment of Lorentz transformation
between inertial reference systems actually needs the invariance of
Minkowski metric or the form invariance of spacetime interval (an absolute
element in the theory) as 
such additional symmetry for the spacetime
structure. In \cite[p.121]{M52}, a procedure is given to determine the form 
of the spacetime
coordinate transformation between two systems in an arbitrary
rectilinear motion relative to each other in Minkowski spacetime,   
but it doesn't hold
good on a more general spacetime manifold for the lack of 
the additional symmetry. Again, from pratical point of view, 
it is concluded that only 
homogeneous spacetimes 
are relevant to the physical covariance under the arbitrary spacetime coordinate
transformation (1).  

\section{ Application to the linear transformation $ \Lambda^{\mu'}_{\mu}$
in homogeneous spacetime (constant
metric fields)}
This is a situation where such important tool as Green's function
method is applicable in dealing with the linear field equations.
The field $A^{\mu}(x)$ and its source field $J^{\mu}(x)$ are then
connected by the Green's function (propagator):
\begin{eqnarray}
A^{\mu}(x)=\int\limits_{M}d^4y~ G(x,y)J^{\mu}(y). 
\end{eqnarray}
Hence our sufficient and necessary 
condition
can be expressed by the first diagram, with the perpendicular 
arrows representing
the combinaton of the construction of Green's function and the performance
of the integral (in Eq. (18)) in system K and K', respectively. 
The requirement for the commutation
of the diagram is therefore translated to the following relation:
\begin{eqnarray}
det(\Lambda ^{\mu'}_{\mu}) G'(x',y')=G(x,y).
\end{eqnarray}

With the help of the verbian fields $e^{\mu}_{a}$,
which connect $g_{\mu \nu}$ and Minkowski metric $\eta_{ab} $ as
\begin{eqnarray}
\eta_{ab}=e^{\mu}_{a}e^{\nu}_{b}g_{\mu \nu},
\end{eqnarray}
we can construct
through Fourier transformation the Green's function (propagator) in 
system K (resp. K'):
\begin{eqnarray}
G(x,y)&=& det(e^{a}_{\mu})\int \frac{(dk_{a})}{(2 \pi)^4}~\frac{1}
{\eta^{ab}k_{a}k_{b}}~e^{ik_{a}e^{a}_{\mu}(x^{\mu}-y^{\mu})}\\
&=&det(e^{a}_{\mu})\frac{1}{4\pi R}\theta (x^0-y^0)
[\delta (e^0_{\mu}(x^{\mu}-y^{\mu})+R)+\delta (e^0_{\mu}(x^{\mu}-y^{\mu})-R)]
,
\end{eqnarray}
where
$R^2=\sum_{i=1}^{3}\left(e_{\mu}^{i}(x^{\mu}-y^{\mu})\right)^2$, and only
the second term contributes because we take $R> 0$.

From the singularity of the $\delta$ function in Eq. (22)
we directly obtain the light cone equation
\begin{eqnarray}
g_{00}(wt)^2+2g_{0i}(wt)x^{i}+g_{ij}x^{i}x^{j}=0, 
\end{eqnarray}
with which the physical propagation speed of interaction $w$ 
(in the sense of the isotropic one in \cite[p.82]{L90} in
the specific system of coordinate can be obtained.
The covariance of Green's function requires 
that $w$ 
should remain invariant under linear spacetime coordinate transformations
\footnote[2] { From the equivalence of K and K', the group parameters
$\beta_{i}(w)$
in
the transformation formulae of tensors from K to K',
e.g. in that of one-order tensor $x^{\mu}$,
$ x^{\mu'}= \Lambda^{\mu'}_{\mu}\left(\beta_{i}(w)\right)x^{\mu},$
should depend on the propagation
speed of interaction $w$ in K,
whereas the transformation group parameters
$\beta_{i}(w')$ of
those from K' to K should depend on the propagation speed of interaction 
$w'$ 
measured in K'. We have
$\Lambda^{\nu}_{\mu'}\left(\beta_{i}(w')\right)
\Lambda^{\mu'}_{\mu}\left(\beta_{i}(w)\right)\not=\delta^{\nu}_{\mu}$, if the
propagation speed of interaction doesn't remain invariant
under spacetime coordinate
transformation. Therefore, Green's function constructed in K' can't be
connected with that constructed in K through Eq. (19), because
$$e^a_{\mu}x^{\mu}
=e^a_{\mu}\Lambda^{\mu}_{\mu'}\left(\beta_{i}(w')\right)x^{\mu'}
\not=e^a_{\mu}\Lambda^{\mu}_{\mu'}\left(\beta_{i}(w)\right)x^{\mu'}
=e^a_{\mu'}x^{\mu'}$$
in the variable of $\delta$ function.}.
The same argument applies to the situation on Riemannian manifold too,
if we consider the covariance of the equation of local lightcone derived
from the local limit of Green's function.
Going back to the real
physical world, Minkowski spacetime, we have the universal
propagation speed $c$ for all kinds of long-range interactions
because of the uniqueness of energy-momentum relation ( the
conjugate of Eq. (23) in momentum space) for the massive particles; 
otherwise in
system K' one of them will be
\begin{eqnarray}
{\bf w'}=\left(\frac{w_1-v}{1-w_1 v/c^2},\frac{w_2\sqrt{1-v^2/c^2}}{1-w_1 v/c^2},
\frac{w_3\sqrt{1-v^2/c^2}}{1-w_1 v/c^2}\right),
\end{eqnarray}
and it breaks the covariance
of the Green's function required by Eq. (19). Thus the invariance of light 
speed is
indispensible as one of the two postulates for the special theory of 
relativity,
since it starts with an attempt at a covariant electromagnetic field theory
between inertial syetems of reference.
It should be specially emphasized that, because the integral range
is over the whole spacetime manifold in Eq. (18), the invariant propagation
speed must exist at every point in space and keep invariant with the
progress of time. 
This requirement imposes the other constraint on the covariance of the
field theory described by Eq. (2).

Finally, we obtain from the above discussion three physical results
related to the propagation speed of interaction in Minkowski spacetime:

1) The invariance of propagation speed of interaction under spacetime
transformation will further limit the form of admissible linear
transformation groups preserving the covariance of the field theory
descibed by Eq. (2). In the case of 1+1 dimension, for example,
the form of such linear transformations:
\begin{eqnarray}
ct'=a_{00}ct+a_{01}x^{1},\\ 
x^{1'}=a_{10}ct+a_{11}x^{1},
\end{eqnarray}
should be given the following constraint:
\begin{eqnarray}
a_{01}=a_{10},\\
a_{00}=a_{11}, \\ 
a_{00}^2-a_{01}^2=1.
\end{eqnarray}
A proper Lorentz transformation obviously falls into the class.

2) To make the situation more complicated we put some refractive media,
which gives rise to a refractive index $n\neq 1$, into the spacetime manifold.
Certainly the density of it is low enough ($T^{00}\ll 1$) not to influence
the spacetime metric too much, but the eletromagnetic propagation speed $c/n$ is
no longer invariant under coordinate transformation in where the media
is distributed. The above discussion tells us that any small patch of such
media can destroy the covariance of electromagetic field, even if the field 
equations are still form covariant
on the whole rest of the spacetime manifold. Thus the covariance of
electromagnetic field is only
an approximate symmetry in reality.

3) If the mass term $ -\frac{1}{2}m^2 A^{\mu}A_{\mu}$ is added to the
Lagrangian
density in Eq. (2), then the contribution to the field $A^{\mu}(x)$ at
point $x$ in Minkowski spacetime comes
from the source field within the backward light cone of the point. With
the time reversal symmetry of light cone, the
covariance of the field requires that the
velocity of the massive quanta of $A^{\mu}(x)$ emitted by its source 
range from $-c$ to $c$, so the
four-momentum of them will diverge. Independently of the requirement for
gauge symmetry, therefore, a massive long-range intermediate Boson
field is forbidden in nature by the requirement for covariance.

\vspace{6mm}
{\bf ACKNOWLEDGMENTS}. I wish to thank Shihai Dong, Guangshun Huang, Lin Zhang
for the extensive helps in completing the paper.

\end{document}